\documentclass[reprint,aps,prl,twocolumn,superscriptaddress,groupedaddress]{revtex4-2}

\usepackage[final]{graphicx}
\usepackage[hidelinks]{hyperref}
\usepackage{bm,physics,amsmath,amssymb,xcolor}
\usepackage[normalem]{ulem}

\usepackage{centernot}
\usepackage{braket}
\usepackage{tabularx}
\usepackage{booktabs}
\usepackage{pifont}
\usepackage{colortbl}
\usepackage{amsthm}
\usepackage{soul}

\usepackage{titlesec}

\newcommand{\eomo}{$E_1$-$M_1$}
\newcommand{\eoet}{$E_1$-$E_2$}
\newcommand{\etet}{$E_2$-$E_2$}

\begin{document}

\title{Bottom-up Analysis of Ro-Vibrational Helical Dichroism}

\author{Mateja Hrast}
\email[]{mateja.hrast@ist.ac.at}
\affiliation{Institute of Science and Technology Austria (ISTA), Am Campus 1, 3400 Klosterneuburg, Austria}
\author{Georgios M. Koutentakis}
\email{georgios.koutentakis@ist.ac.at}
\affiliation{Institute of Science and Technology Austria (ISTA), Am Campus 1, 3400 Klosterneuburg, Austria}
\author{Mikhail Maslov}
\email{mikhail.maslov@ist.ac.at}
\affiliation{Institute of Science and Technology Austria (ISTA), Am Campus 1, 3400 Klosterneuburg, Austria}
\author{Mikhail Lemeshko}
\email{mikhail.lemeshko@ist.ac.at}
\affiliation{Institute of Science and Technology Austria (ISTA), Am Campus 1, 3400 Klosterneuburg, Austria}

\date{\today}

\begin{abstract}

We present a general theoretical framework for helical dichroism (HD), establishing an explicit link between chiral resolution and orbital angular momentum (OAM) exchange in light–matter interaction. Tracing microscopic mechanisms of the OAM transfer, we derive rotational selection rules, which establish that HD emerges only from the spin–orbit coupling of light, even for beams without the far-field OAM. Our findings refine the conditions for observing HD, provide a tool to re-examine the outcome of prior experiments, and guide future designs for chiral sensing with structured light.

\end{abstract}

\maketitle

Many biologically and chemically important molecules exist as chiral pairs -- two non-superimposable mirror versions called enantiomers. The ability to separate the enantiomers is essential for the pharmaceutical and agrochemical industries \cite{MAIER2001}. This is perhaps best illustrated by the methamphetamine molecule, whose R-enantiomer is an effective nasal decongestant, while its S-enantiomer is a psychoactive drug that fuels addiction epidemics~\cite{barkholtz2023}. The high economic relevance has driven research on the asymmetric synthesis of chiral molecules,~e.g.,~via an enantioselective or bio-catalysis or by employing chiral auxiliaries \cite{Brown1989}. However, these advanced synthesis techniques require further refinement and the assessment of enantiomeric purity by enantio-selective processes. This next step, while essential, is still technically challenging. Existing chemical techniques each present their own challenges, ranging from high cost and time consumption to limited applicability and reliability \cite{qian2023}. Here, we address an alternative, chemical-physics perspective, aiming to assess enantiomeric purity through light–matter interactions. Owing to rapid advances in optical physics \cite{Koch2019}, this approach promises a cost-effective and highly controllable avenue for chiral discrimination.

For decades, the difference in absorption of left- and right-circularly polarised light, known as the circular dichroism (CD) \cite{deutsche1970,Holzwarth1974}, has been used in spectroscopy of chiral media \cite{Miles2021}. However, circular dichroism relies on a limited resource, namely, the spin angular momentum of light. Meanwhile, the pioneering study by Allen~\textit{et~al.}~\cite{Allen1992} has demonstrated that, in addition to spin, light can also carry orbital angular momentum (OAM). In contrast to spin, the OAM of a beam is, in principle, unbounded. The helical wavefronts of twisted light, also known as optical vortex beams, fascinated physicists, who envisaged that their chiral structure could match the chirality of matter, offering unprecedented possibilities for enantioselective applications~\cite{Andrews2021}.

This fascination culminated in various concrete OAM-induced processes being proposed as probes of enantiomeric purity, particularly, in the form of helical dichroism (HD)~\cite{ANDREWS2004,Ye2019,Li2021}, which is argued to represent a significant improvement over CD \cite{Ye2019,Li2021}. Theoretical works of Forbes and Andrews \cite{Forbes2018,Forbes2019,Forbes2021} laid the groundwork for HD, by carefully analyzing the symmetry properties of the terms in the multipolar expansion of light-matter interaction, and pinpointing the electric-dipole–magnetic-dipole (\eomo), electric-dipole–electric-quadrupole (\eoet) and electric-quadrupole–electric-quadrupole (\etet) terms as chiral discriminators. These authors derived the expression for the \eoet ~transition amplitude in terms of molecular (transition) multipole moments; however, they made no attempt to evaluate the latter in terms of molecular rotational states. On the experimental side, initial attempts showed no signs of HD \cite{Araoka2005,Loeffler2011}; however, recent experiments have claimed to observe it \cite{Rusak2019,Zhang2020,Rouxel2022,Begin2023,Jain2023}. Nonetheless, the experimental apparatus of these experiments is very complex, and, although it was argued that tight focusing of a beam is a requirement \cite{Forbes2019}, so far there have been no theoretical attempts to pinpoint the origin of the observed dichroism. 

Clearly, a microscopic theory is needed that (i) links the observed dichroic signals to contributing ro-vibrational transitions and the associated transfer of different types of angular momenta, and (ii) identifies the required molecular symmetries or focusing conditions that enable HD. This work fills this gap in the literature by establishing an explicit connection between chiral resolution and orbital angular momentum exchange in light-molecule interaction, and identifying the conditions that give rise to HD in ro-vibrational spectroscopy, paving the way for further advancements of this field.

We first provide an overview of chirality detection methods based on molecular multipole moments, and derive symmetry requirements for observing \eoet-based dichroism. Using our recently introduced molecule-light interaction Hamiltonian \cite{Maslov2024,Maslov_Thesis}, we derive the selection rules for ro-vibrational transitions induced by the \eoet~term. By tracing the angular momentum transfer involved in these rules, we demonstrate that paraxial Laguerre-Gaussian (LG) beams cannot produce the \textit{genuine} HD,~i.e.,~HD involving the transfer of OAM from light to the molecule. Instead, we show that tight-focusing associated with the spin-orbit coupling (SOC) of light~\cite{Bliokh2015} enables OAM transfer to chiral molecules by introducing HD contributions that are absent within the paraxial approximation. Strikingly, even a Gaussian beam, which carries no OAM in the far-field, can produce a HD signal in the tightly-focused regime. Collectively, our findings refine the theoretical foundations of HD in ro-vibrational spectroscopy, and guide the designs of future experiments and enantiomer-sensitive techniques using structured light.

Existing enantiomer-sensitive spectroscopy techniques rely on the analysis of the multipole moments of the molecule. The (vibrational) circular dichroism---the standard method to detect chirality---is based on the \eomo~coupling and has already been exhaustively investigated \cite{Stephens1985,BUCKINGHAM1987,Mun2019,Lovesey2019}. It draws on the fact that the electric dipole moment of a chiral molecule transforms as a vector, and thus picks up a minus sign, under mirror reflection, while the magnetic dipole, which is a pseudovector, is invariant to it. A coexistence of both these multipoles implies that a molecule lacks a mirror symmetry plane (strictly speaking, an improper rotation axis) and is, by definition, chiral. This type of CD can be exhibited by  chiral molecules, for which the symmetry point group permits at least one nonzero dipole moment. However, the dichroism stemming from the \eomo ~term only depends on the intensity of the field, which makes other sources, where the dichroism can be enhanced by the structure of the field, attractive alternatives.

One example\textemdash the phase sensitive microwave three-wave mixing technique \cite{Patterson2013PRL,Patterson2013}\textemdash exploits the fact that a non-zero product of all dipole moments $d_xd_yd_z\neq 0$ implies that the molecule is chiral~\cite{Patterson2013,Ordonez2018,Ayuso2022}. Symmetry analysis reveals that only completely asymmetric ($C_1$) chiral molecules can be characterized by this condition.

Our work explores another direction -- the conditions for chirality stemming from the electric dipole and quadrupole moments of the molecule. The dichroism induced by the \eoet~term can be enhanced with appropriately structured fields, which makes it a promising prospect for future enantiomer-sensitive techniques. Recently, the magnetic-dipole-magnetic quadrupole ($M_1$-$M_2$) coupling has also been suggested as a source of HD, which is equivalent to the \eoet~term from a symmetry standpoint \cite{Ji2024}. However, this and other higher-order terms, such as \etet, are difficult to detect experimentally due to low transition probabilities in the multipole expansion.

The remainder of this paper focuses on characterizing the dichroism originating from the \eoet ~term. Using a simple point-charge model, we derive the following condition (proof in the Supplementary Material~\footnote{See Supplementary Material, containing references \cite{Maslov2024,Maslov_Thesis,Lax1975,Bliokh2015,Bliokh2023}}):\\

\textit{If the product of dipole and quadrupole moments $d_{\mu}Q_{\nu \rho} \neq 0$ for distinct coordinates $\mu, \nu, \rho \in \{x,y,z\}$ of the principal axis system, then the molecule is chiral.} \\

To date, the above condition has always been associated with HD~\cite{ANDREWS2004,Forbes2018}. However, as we show in this work, it can also lead to discriminatory effects without an OAM transfer from light to the internal molecular degrees of freedom. To resolve this ambiguity, this work uses \textit{``helical dichroism''} to denote dichroism effects where \emph{OAM transfer occurs}, while dichroism without such transfer, even for an OAM-carrying beam, is referred to as CD.

Table~\ref{tab:chiral_multipole_dofs} summarizes which chiral point groups can exhibit \eoet-induced dichroism. In the case of pure-rotational spectroscopy, only the permanent multipole moments of the molecule are relevant. In ro-vibrational spectroscopy vibrational transitions can alter the symmetry of a molecular state. Therefore, transformation properties of vibrational states in terms of their associated irreducible representation $V$ need to be considered, which allows for the detection of more chiral groups. Table~\ref{tab:chiral_multipole_dofs} provides the list of $V$'s exhibiting dichroism for each chiral point group.

 \begin{table}[ht!]
    \centering
    \caption{The \eoet ~chiral signatures in rotational and ro-vibrational spectroscopy of all possible chiral point groups. In the case of ro-vibrational spectroscopy, we provide the irreducible-representations of the vibrational modes associated with non-zero chiral signature.}
     \setlength\tabcolsep{3pt}
\begin{tabular}{l | c c c c c c c c c c}
\toprule
     Point group     & $C_1$ & $C_2$ & $C_{n\geq 3}$ & $D_2$ & $D_{n\geq 3}$ & $T$ & $O$ & $I$ \\ \midrule
     Rotational      & \textcolor{black}{\ding{52}} & \textcolor{black}{\ding{52}}& \textcolor{red}{\ding{56}}  & \textcolor{red}{\ding{56}}  & \textcolor{red}{\ding{56}}  & \textcolor{red}{\ding{56}}  & \textcolor{red}{\ding{56}}  & \textcolor{red}{\ding{56}} \\ 
     Ro-vibrational  & $A$ & $A$, $B$ & $E_1$ & $B_{1, 2, 3}$ & $E_1$ & $T$ & \textcolor{red}{\ding{56}} & \textcolor{red}{\ding{56}} \\  
     \bottomrule
\end{tabular}
     \label{tab:chiral_multipole_dofs}
 \end{table}

As displayed in Table \ref{tab:chiral_multipole_dofs} the \eoet~term can only identify chiral molecules with $C_1$ and $C_2$ symmetries via transitions that preserve the molecular symmetry (pure rotational and ro-vibrational of the trivial representation $A$). Since most known chiral molecules are $C_1$-symmetric,~i.e.,~they are asymmetric tops \cite{Bernath}, we restrict our scope to these molecules.

Eigenstates of the asymmetric top rotor Hamiltonian, expressed in the standard rotational basis $\ket{J,M,K}$~\cite{Bernath}, read 
\begin{equation}
    \ket{J_{N_p,N_o}(M)}=\sum_KC_K\ket{J,M,K}.
    \label{basisChange}
\end{equation}
Here, $J$ is total angular momentum quantum number, $M$ and $K$ are the projections of ${\bm J}$ on the lab quantization axis $z$, herewith fixed to the propagation axis of light, and the body-fixed principal axis respectively. The asymetric top rotor indices $N_p$ and $N_o$ are the limits of $K$ in the pure prolate/oblate solutions and are not good quantum numbers. Since quantum numbers $J$ and $M$ are not affected by the change in basis \eqref{basisChange}, we proceed to derive the selection rules for the electric dipole ($E_1$) and quadrupole ($E_2$) transitions within the rotational basis, $| J, M, K \rangle$.

The transition rate from an initial $i=\ket{J,M,K}$ to a final $f=\ket{J,M',K'}$ molecular rotational state can be calculated using the light-matter interaction Hamiltonian, derived in Ref. \cite{Maslov2024,Maslov_Thesis}. The transition amplitude reads
\begin{equation}
    \mathcal{M}_{i\to f}=\sum_{lm\mu}\mathcal{I}^{\text{vib}; l,\mu}\,\mathcal{I}^{\text{CM}; l,m}\,\mathcal{I}^{\text{rot}; l,m,\mu,\sigma}_{J,M,K,J,M',K'}\,,
    \label{eq_transition_matrix}
\end{equation}
where $l=0$ and $l=1$ for dipole and quadrupole transitions respectively, $\mu = 0, \pm 1, \dots, \pm (l+1)$ runs over all components of a given multipole moment in the molecule-fixed frame, while both $m = 0, \pm 1, \dots, \pm l$, the OAM transferred to the molecule, and $\sigma =0, \pm 1$, the polarization component of light, correspond to lab-frame angular momenta. The vibrational, $\mathcal{I}^{\text{vib}; l,\mu}$, and center-of-mass, $\mathcal{I}^{\text{CM}; l,m}$, terms do not depend on molecular rotational states. Hence, in this work, we focus on the rotational transition amplitude $\mathcal{I}^{\text{rot};l,m,\mu,\sigma}_{J,M,K,J,M',K'}$,  which is proportional to the Clebsh-Gordan coefficient $C^{J, M}_{j_1, m_1, j_2, m_2}$~\cite{Maslov2024,Maslov_Thesis}. From this coefficient, we derive the angular momentum selection rules, summarized in Table~\ref{SelectionRules}~\cite{Note1}.

\begin{table}[ht!]
    \caption{Selection rules for electric dipole ($E_1$) and quadrupole ($E_2$) transitions between rotational states $\ket{J,M,K}$ and $\ket{J',M',K'}$. {For details on forbidden transitions, see \cite{Note1}.}}
    \centering
    \begin{tabular}{c c}\hline
    \toprule
        \textbf{$E_1$ transition} & \textbf{$E_2$ transition}  \\
        \midrule
        $\Delta J\leq 1$ &  $\Delta J\leq 2$ \\
        $\Delta M=-\sigma$ & $\Delta M=m-\sigma$ \\
        $\Delta K=\pm\mu$ & $\Delta K=\pm\mu$ \\
        $J=0\not\to J'=0$ &  (see \cite{Note1}\\
        $\ket{J,0,K}\not\to\ket{J,0,K'}$ &  for details)\\
    \bottomrule
\hline
    \end{tabular}
    \label{SelectionRules}
\end{table}

To realize an \eoet ~transition, both $E_1$ ($l=m=0$) and $E_2$ ($l = 1$, $m = 0, \pm 1$) transitions should be allowed between the chosen two states simultaneously. According to the $\Delta M$ selection rules (see Table~\ref{SelectionRules}), for light with a well-defined polarization, $\sigma$, this is only possible for the $m = 0$ component of the beam, which carries no OAM. This brings us to the first key finding of our work: The dichroism stemming from the \eoet~term of a paraxial beam should not be associated with HD, as no OAM transfer to the molecule occurs. Moreover, this dichroism is present even in the case of a plane-wave and thus is not associated at all to the structure of light or its helicity. 

This raises the question whether the \eoet~term can produce HD, where OAM transfer occurs. In what follows, we show that we can create such a scenario in the case of non-paraxial beams.

\begin{figure}[ht!]
    \centering
    \includegraphics[width=\linewidth]{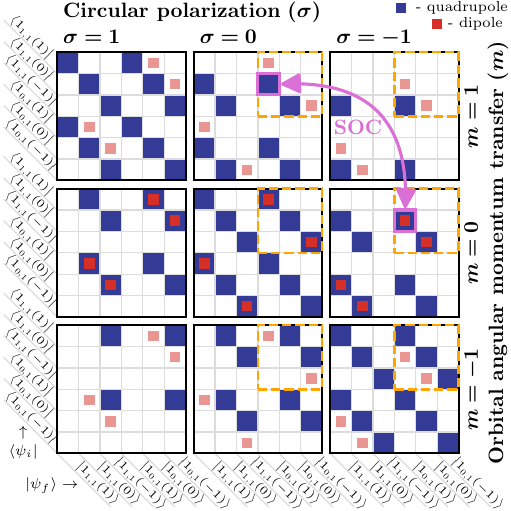}
    \caption{Allowed dipole (red squares) and quadrupole transitions (blue squares) between eigenstates of the asymmetric top Hamiltonian. Panel columns represent different circular polarizations of the beam, and panel rows -- the amount of the OAM that can be transferred from the beam to the internal rotation of the molecule. Purple circles show an example of SOC creating an \eoet ~term with the transfer of OAM. Orange squares highlight the transitions that would be superimposed in the experiment, as shown in Fig. \ref{fig:profiles}.}
    \label{fig:gridJ1}
\end{figure}

In Fig.~\ref{fig:gridJ1}, we demonstrate the allowed dipole (red squares) and quadrupole (blue squares) transitions between the rotational eigenstates of asymmetric top molecules. For clarity, only a selection of eigenstates is shown here. The full figure up to $J\leq 2$ can be found in Ref.~\cite{Note1}, but it provides the same qualitative picture.

The three columns of Fig.~\ref{fig:gridJ1} correspond to different beam polarizations $\sigma$, and the three rows to different OAM transfer quantum numbers $m$. As shown in Ref.~\cite{Maslov2024,Maslov_Thesis}, electric dipole transitions do not involve the OAM transfer, so they appear only in the $m=0$ row (though we also display them in the $m=\pm 1$ rows for comparison). In contrast, electric quadrupole transitions can involve the transfer of up to one OAM quantum ($m=0,\pm 1$). Any non-planar beam profile generally contains contributions in all three rows, but their relative weights depend on the specific beam structure.

We first focus on the red-blue squares, indicating pairs of states connected by both dipole and quadrupole transitions. All of these appear in the $m=0$ row, which denotes transitions where no OAM is transferred. This confirms the findings of Table~\ref{SelectionRules} and suggests that for beams with well-defined $\sigma$ the \eoet ~term cannot create HD, but is merely an additional contribution to CD.

However, the clear separation of the OAM in terms of~$m$, and spin in terms of $\sigma$, is only valid for paraxial beams. In fact, any kind of SOC of light would result in a coupling of different $\sigma$ components~\cite{Bliokh2015,Bliokh2023}. For our purposes, this means that the terms from different columns in Fig.~\ref{fig:gridJ1} are enabled to drive the same transition. This would, for example, give rise to the \eoet ~term corresponding to the purple square with a transition that is simultaneously dipole (at $m=0$, $\sigma=-1$) and quadrupole (at both $m=0$, $\sigma=-1$ and $m=1$, $\sigma=0$): $| 1_{0,1}(1) \rangle \to | 1_{1,1}(0) \rangle$. The coupling of the dipole moment term to the quadrupole component with $m = -1$ enables the molecule to absorb the OAM of light since $m \neq 0$.

\begin{figure}[t!]
    \centering
    \includegraphics[width=\linewidth]{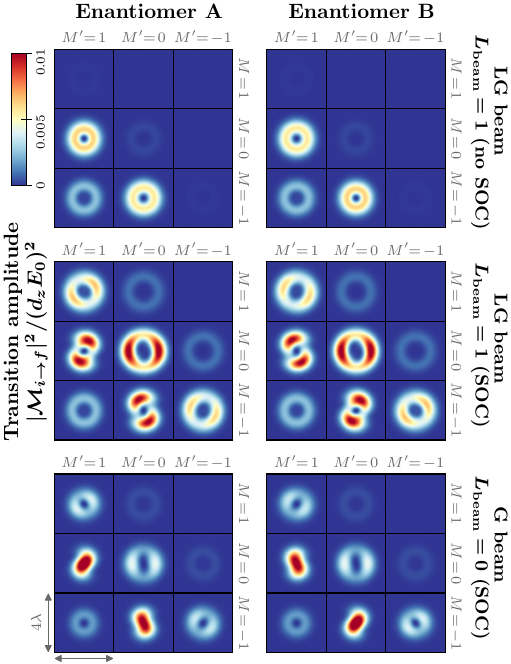}
    \caption{Signal profiles of the \eoet ~transitions from $\bra{1_{1,1}(M)}$ to $\ket{1_{0,1}(M')}$ for the opposite enantiomers coupled by a Laguerre Gaussian beam without (top) and with (middle row) the SOC and by a Gaussian beam with SOC (bottom).}
    \label{fig:profiles}
\end{figure}

The SOC of light, required to realize such transitions, can be achieved,~e.g.,~by the tight focusing of  Laguerre-Gaussian beams \cite{Loeffler2011,Forbes2021nonparaxial,Forbes2021longitudinal}. The electric field at the focus of the beam (focal field) induces the interconversion between the spin and OAM, manifesting as the SOC of light \cite{Note1,Bliokh2015,Bliokh2023}. Here, we describe the beam in the first-order non-paraxial approximation \cite{Note1, Lax1975}, and use the interaction Hamiltonian of Ref.~\cite{Maslov2024,Maslov_Thesis} to calculate the spatially-resolved matrix elements between the states denoted by a dashed square in Fig.~\ref{fig:gridJ1}. The chosen parameters are: transition quadrupole moment $Q_{xy}=0.5\,d_z\lambda$, where $d_z$ is the transition dipole moment and $\lambda$ the wavelength, focal beam-waist $\omega_0=0.2\,\lambda$, distance from focus $z=\lambda$, with $z$ being the propagation direction, azimuthal charge of the Laguerre-Gausian $L_{\rm beam}=1$ and polarization $\sigma=-1$. Note that changing $L_{\rm beam} \to - L_{\rm beam}$ would only affect the relative amplitudes of the $M'=-1 \to M=1$ and $M'= M =0,\pm 1$ transitions, and not the spatial distributions. The opposite enantiomers are described by reflection on the molecule fixed (inertial) $xy$-plane. It is well-enstablished \cite{Buckingham1971, Power1975} that an anisotropy is needed to observe the \eoet~ interaction, which is why $M$ state resolution is required in any proposed HD experiment. The obtained signal profiles for states with resolved $M$ are shown in Fig.~\ref{fig:profiles}.

We first note that the integral of all shown profiles is equal, unless $d_zQ_{xy}^*$ has an imaginary component ${\rm Im}[d_zQ_{xy}^*]$, which leads to dichroism at the level of total absorption. This is a known source of vibrational CD, and is independent of the structure of the electric field and only depends on the internal parameters of the molecule, see \cite{Buckingham1971}. From this result we conclude that HD is impossible to detect by just the magnitude of the absorbed intensity of light, but requires the spatial resolution of the absorption profile~\cite{Loeffler2011}.

Without the light SOC, the only difference between the two enantiomers is the intensity of the $M=0\to M'=1$ and $M=-1\to M'=0$ profiles, see top row of Fig.~\ref{fig:profiles}. This dichroism originates from the $m=0$ part of the beam -- the red-blue squares within the dashed square in the $\sigma=-1$, $m=0$ panel of Fig.~\ref{fig:gridJ1}. Therefore, this is another source of spatially-resolved vibrational CD, as no OAM is transferred. 

Tight focusing introduces SOC of light, which gives rise to new distinguishing effects between the enantiomers in the absorption profiles, seen in the middle row of Fig.~\ref{fig:profiles}. First, there is a large difference in the spatially-resolved absorption of $M=-1\to M'=0$ and $M=0\to M'=1$ transitions. The signal comes from the coupling of dipole and quadrupole transitions in the dashed squares in Fig.~\ref{fig:gridJ1}. This difference stems from the relative handedness of the vortex beam and the chiral molecule and involves transfer of OAM. Therefore, this can be considered genuine HD.

Next, there is also a difference in the $M=1\to M'=1$ and $M=-1\to M'=-1$ profiles. It arises from the coupling of dipole transitions in the dashed squares with $\sigma=0$ (Fig.~\ref{fig:gridJ1}) to the quadrupole transitions in the $m=-1$, $\sigma=-1$ square. Therefore, this is also HD. Overall, the dichroism signal in the case of SOC light is an order of magnitude stronger than without the coupling.

The bottom row in Fig. \ref{fig:profiles} shows that in the case of SOC, it is not necessary to use a beam with OAM in the asymptotic limit. Indeed, HD is also observed in the case of a tightly-focused Gaussian beam. This is because the focal field, through the conversion of spin into OAM, takes the form of an LG mode with the azimuthal charge~$\sigma$. In fact, a similar dichroism is also seen for a tightly-focused linearly polarized beam, because the two components of circular polarization within it couple through the SOC. This shows that neither OAM nor spin of the incoming beam are an essential component to observe HD, but rather the SOC.

Note, that due to the SOC of light, the type of angular momentum transferred to the molecule cannot be easily distinguished, which hinders the characterization of the resulting dichroism as either CD or HD. Our work pinpoints the OAM and spin absorption channels in terms of the quantum numbers $m$ and $\sigma$ (see Fig.~\ref{fig:profiles}) providing a way to resolve this issue.

In summary, we have shown that paraxial vortex beams cannot produce HD, as any observed dichroism under these conditions reduces to the standard CD. Genuine HD---dichroism directly tied to the transfer of OAM---requires SOC of light, as in the case of tightly-focused beams. Strikingly, a purely Gaussian mode can exhibit genuine HD once sufficiently focused, even though it carries no OAM in the far-field. These findings clarify the origin of HD in structured-light experiments and suggest that prior measurements of HD may warrant a careful re-examination to confirm the presence of OAM transfer. In particular, the studies that did not observe dichroism were either probing primarily dipole-order interactions \cite{Araoka2005} or suppressed the dichroism signal due to averaging \cite{Loeffler2011}. The studies that successfully detected dichroism, on the other hand, employed both quadrupole field interactions and tightly-focused beams \cite{Rusak2019,Rouxel2022,Begin2023,Jain2023}. To properly model the full complexity of these experiments, the here-presented theory should in the future be expanded to include the electronic degrees of freedom. This will allow us to address processes that involve electronic transitions, including complex multi-photon processes, and explain new emerging experiments in both atoms and molecules~\cite{Begin2025}.

In light of our results, the experimental setup for ro-vibrational chiral-sensitive spectroscopy and enantiomer separation that leverages the spatial structure of light is expected to be complex and demanding~\cite{Leibscher2022}. Nevertheless, the framework developed here offers a roadmap for future experimental designs and a versatile tool to quantify both HD and the newly identified CD signatures.\\

\begin{acknowledgments}
This research was funded in whole or in part by the Austrian Science Fund (FWF) [10.55776/F1004]. \cite{data}
\end{acknowledgments}

\bibliography{bibliography}

\begin{thebibliography}{46}
\bibitem [{dat()}]{data}%
  \BibitemOpen
  \href@noop {} {}\bibinfo {note} {The data that support the findings of this
  article are openly available at 10.5281/zenodo.17864838}\BibitemShut
  {NoStop}%
\end{thebibliography}

\end{document}